\documentclass[10pt,conference]{IEEEtran}
\IEEEoverridecommandlockouts
\usepackage{graphicx}
\usepackage[table]{xcolor}

\usepackage{cite}
\usepackage{glossaries}
\usepackage[hidelinks, bookmarks=false]{hyperref}
\usepackage[noabbrev,capitalise]{cleveref}
\usepackage{booktabs}
\usepackage{array}
\usepackage[symbol]{footmisc}
\usepackage{threeparttable}
\usepackage{makecell}
\usepackage{arydshln}
\usepackage{tikz}
\usepackage{eso-pic}
\usepackage{hyphenat}
\usepackage{listings}
\usepackage{microtype}
\usepackage{amsmath,amsfonts}
\usepackage{soul}
\usepackage{algorithm}
\usepackage{algpseudocode}
\usepackage{enumitem}
\usepackage{multirow}
\usepackage{xspace}
\usepackage{pifont}
\usepackage{changepage}
\usepackage{multicol}
\usepackage{amssymb}
\usepackage{printlen}
\usepackage{newtxtext,newtxmath}

\makeatletter
\newcommand\fs@spaceruled{%
  \def\@fs@cfont{\bfseries}%
  \let\@fs@capt\floatc@ruled%
  \def\@fs@pre{\vspace{.7\baselineskip}\hrule height.8pt depth0pt \kern2pt}%
  \def\@fs@post{\kern2pt\hrule\relax}%
  \def\@fs@mid{\kern2pt\hrule\kern2pt}%
  \let\@fs@iftopcapt\iftrue}
\makeatother

\usepackage[per-mode=repeated-symbol, free-standing-units, allow-number-unit-breaks]{siunitx}
\DeclareSIUnit\comp{COMP}
\DeclareSIUnit\flop{FLOP}
\DeclareSIUnit\flops{FLOPS}
\DeclareSIUnit\bps{bps}
\DeclareSIUnit\Bps{Bps}
\DeclareSIUnit\gate{GE}
\DeclareSIUnit\op{OP}
\DeclareSIUnit\macu{MACU}
\DeclareSIUnit\ops{OPS}
\DeclareSIUnit\core{core}
\DeclareSIUnit\request{request}
\DeclareSIUnit\cycle{cycle}
\DeclareSIUnit\teraops{TOPS}
\DeclareSIUnit\ghz{GHz}
\DeclareSIUnit\mhz{MHz}
\DeclareSIUnit[number-unit-product = ]\percent{\%}

\definecolor{MidnightBlue}{HTML}{191970}
\definecolor{Mint}{HTML}{3EB889}
\definecolor{EnglishRed}{HTML}{A4515C}
\definecolor{SelectiveYellow}{HTML}{FFBA08}
\definecolor{CyanProcess}{HTML}{08B2E3}
\definecolor{OliveDrab7}{HTML}{4D4730}
\definecolor{Red}{HTML}{FF0000}
\colorlet{color1}{MidnightBlue}
\colorlet{color2}{Mint}
\colorlet{color3}{EnglishRed}
\colorlet{color4}{SelectiveYellow}
\colorlet{color5}{CyanProcess}
\colorlet{color6}{OliveDrab7}
\colorlet{colorAlert}{Red}
\definecolor{PulpGreen}{HTML}{168638}
\definecolor{PulpBlue}{HTML}{1269b0}
\definecolor{PulpRed}{HTML}{a8322c}
\definecolor{PulpYellow}{HTML}{f2c100}

\Crefname{equation}{Eq.}{Eqs.}
\Crefname{figure}{Fig.}{Figs.}
\Crefname{tabular}{Tab.}{Tabs.}

\newacronym{ieee}{IEEE}{IEEE}
\glsunset{ieee}

\makeatletter \newcommand{\AddSpaceIfAnonymous}{\@ifclasswith{acmart}{anonymous}{\vspace{10mm}}{}} \makeatother
\PackageWarning{TODO:}{#1!}

\def\BibTeX{{\rm B\kern-.05em{\sc i\kern-.025em b}\kern-.08em
    T\kern-.1667em\lower.7ex\hbox{E}\kern-.125emX}}

\newacronym{ai}{AI}{Artificial Intelligence}
\newacronym{ml}{ML}{Machine Learning}
\newacronym{nn}{NN}{Neural Network}
\newacronym{cpu}{CPU}{Central Processing Unit}
\newacronym{asic}{ASIC}{Application Specific Integrated Circuit}
\newacronym[longplural={Systems-on-Chip}]{soc}{SoC}{System-on-Chip}
\newacronym{fpga}{FPGA}{Field Programmable Gate Array}
\newacronym{asip}{ASIP}{Application Specific Instruction Processor}
\newacronym{gpp}{GPP}{General Purpose Processor}
\newacronym{gp}{GP}{general-purpose}
\newacronym{gpgpu}{GP-GPU}{General Purpose Graphics Processing Unit}
\newacronym{gpu}{GPU}{Graphics Processing Unit}
\newacronym{sm}{SM}{Streaming Multiprocessor}
\newacronym{cuda}{CUDA}{Compute Unified Device Architecture}
\newacronym{mpi}{MPI}{Message Passing Interface}
\newacronym{cots}{COTS}{Commercial-Off-The-Shelf}
\newacronym{soa}{SoA}{state-of-the-art}
\newacronym{roi}{ROI}{Return on Investments}
\newacronym
[
  longplural={Core Complexes}
]
{cc}{CC}{Core Complex}

\newacronym{lte}{LTE}{Long Term Evolution}
\newacronym{nr}{NR}{New Radio}
\newacronym{4g}{4G}{4th Generation}
\newacronym{5g}{5G}{5th Generation}
\newacronym{b5g}{B5G}{Beyond-5G}
\newacronym{6g}{6G}{6th Generation}
\newacronym{urll}{URLL}{Ultra-Reliable Low-Latency}
\newacronym{mmtc}{mMTC}{massive Machine Type Communications}
\newacronym{embb}{eMBB}{enhanced Mobile Broadband}
\newacronym{3gpp}{3GPP}{3rd Generation Partnership Project}
\newacronym{oran}{O-RAN}{Open-RAN}
\newacronym{ran}{RAN}{Radio Access Networks}
\newacronym{cran}{C-RAN}{Cloud Radio Access Networks}
\newacronym{gnb}{gNB}{Next Generation Node B}
\newacronym{pusch}{PUSCH}{Physical Uplink Shared Channel}
\newacronym{sdr}{SDR}{Software Defined Radio}
\newacronym{phy}{PHY}{Physical}
\newacronym{cu}{CU}{Centralized Unit}
\newacronym{du}{DU}{Distributed Unit}
\newacronym{ru}{RU}{Remote Unit}
\newacronym{ue}{UE}{User Equipment}
\newacronym{ofdm}{OFDM}{Orthogonal Frequency Division Multiplexing}
\newacronym{ofdma}{OFDMA}{Orthogonal Frequency Division Multiple Access}
\newacronym{bf}{BF}{Beam Forming}
\newacronym{mimo}{MIMO}{Multiple-Input, Multiple-Output}
\newacronym{che}{CHE}{Channel Estimation}
\newacronym{dmrs}{DMRS}{Demodulation Reference Signal}
\newacronym{tti}{TTI}{Transmission Time Interval}
\newacronym{sc}{SC}{sub-carrier}
\newacronym{mu}{MU}{Multiple-User}
\newacronym{snr}{SNR}{Signal-to-Noise Ratio}
\newacronym{ber}{BER}{Bit-Error-Rate}

\newacronym{add}{add}{Add}
\newacronym{mul}{mul}{Multiply}
\newacronym{mac}{MAC}{Multiply\&Accumulate}
\newacronym{pmac}{p.mac}{Post-increment Multiply-accumulate}
\newacronym{axpy}{AXPY}{A Times X Plus Y}
\newacronym{dotp}{DOTP}{Dot Product}
\newacronym{sdotp}{SDOTP}{Sum Dot Product}
\newacronym{matmul}{MatMul}{Matrix Multiplication}
\newacronym{gemm}{GEMM}{General Matrix Multiplication}
\newacronym{mvm}{MVM}{Matrix-Vector Multiplication}
\newacronym{fft}{FFT}{Fast Fourier Transform}
\newacronym{sysinv}{SysInv}{Linear System Inversion}
\newacronym{choldec}{CholDec}{Cholesky Decomposition}
\newacronym{mmse}{MMSE}{Minimum Mean Squared Error}
\newacronym{conv2D}{Conv2D}{2D-Convolution}
\newacronym{dct}{DCT}{Direct Cosine Transform}

\newacronym{sram}{SRAM}{Static Random-Access Memory}
\newacronym{dram}{DRAM}{Dynamic Random-Access Memory}
\newacronym{spm}{SPM}{Scratchpad Memory}
\newacronym{tcdm}{TCDM}{Tightly Coupled Data Memory}
\newacronym{IDol}{I\$}{Instruction Cache}
\newacronym{dma}{DMA}{Direct Memory Access}
\newacronym{axi}{AXI}{Advanced eXtensible Interface}
\newacronym{noc}{NoC}{Nework on Chip}
\newacronym{csr}{CSR}{Control Status Register}
\newacronym{hbm}{HBM2E}{High Bandwidth Memory}

\newacronym{ipc}{IPC}{instructions-per-cycle}
\newacronym{wfi}{WFI}{wait-for-interrupt}
\newacronym{raw}{RAW}{read-after-write}
\newacronym{ins}{INS}{instruction}

\newacronym{fpu}{FPU}{Floating Point Unit}
\newacronym{fpss}{FP-SS}{Floating Point Sub-System}
\newacronym{ipu}{IPU}{Integer Processing Unit}
\newacronym{divsqrt}{DIVSQRT}{Division and Square-Root Unit}
\newacronym{lsu}{LSU}{Load Store Unit}
\newacronym{dsp}{DSP}{Digital Signal Processing}
\newacronym{qlr}{QLR}{Queue-Linked Register}

\newacronym{eda}{EDA}{Electronic Design Automation}
\newacronym{ge}{GE}{Gate Equivalent}
\newacronym{fo4}{FO4}{Fan-Out-of-4}
\newacronym{beol}{BEOL}{Back-End-of-Line}
\newacronym{pnr}{PnR}{Place and Route}
\newacronym{ppa}{PPA}{Power, Performance and Area}

\newacronym{numa}{NUMA}{Non-Uniform Memory Access}
\newacronym{fc}{FC}{Fully-Connected}
\newacronym{isa}{ISA}{Instruction Set Architecture}
\newacronym{simd}{SIMD}{Single Instruction Multiple Data}
\newacronym{spmd}{SPMD}{Single Program Multiple Data}
\newacronym{cdf}{CDF}{Cumulative Distribution Function}
\newacronym{api}{API}{Application Programmable Interface}
\newacronym{rtl}{RTL}{Register Transfer Level}
\newacronym{sfr}{SFR}{Synchronization Free Region}
\newacronym{dsl}{DSL}{Domain-Specific Language}

\newacronym{int}{INT}{integer}
\newacronym{fp}{FP}{floating-point}

\newacronym{pe}{PE}{Positional Encoding}
\newacronym{rg}{RG}{Resource Grid}
\newacronym{re}{RE}{Resource Element}
\newacronym{cp}{CP}{Cyclic Prefix}
\newacronym{llr}{LLR}{Log-Likelihood Ratio}
\newacronym{lmmse}{LMMSE}{Linear Minimum Mean Squared Error}
\newacronym{3d}{3D}{3-Dimensional}
\newacronym{2d}{2D}{2-Dimensional}
\newacronym{prb}{PRB}{Physical Resource Block}
\newacronym{mdx}{MDX}{Model Driven Neural Receiver}
\newacronym{dals}{DA-LS}{Data Aided Least Squares}
\newacronym{pals}{PA-LS}{Pilot Aided Least Squares}
\newacronym{cdm}{CDM}{Code-Division Multiplexing}
\newacronym{ldpc}{LDPC}{Low Density Parity Check Code}
\newacronym{bce}{BCE}{Binary Cross-Entropy}
\newacronym{mse}{MSE}{Mean Squared Error}
\newacronym{adam}{ADAM}{Adaptive Moment Estimation}
\newacronym{umi}{UMi}{Urban Microcell}
\newacronym{mcs}{MCS}{Modulation Coding Scheme}
\newacronym{tdl}{TDL}{Tapped Delay Line}
\newacronym{ls}{LS}{Least Squares}
\newacronym{tbler}{TBLER}{Transport Block Error Rate}
\newacronym{cscs}{CSCS}{Swiss National Supercomputing Centre}
\newacronym{flop}{FLOP}{Floating Point Operation}
\newacronym{resblock}{ResBlock}{ResNet block}
\newacronym{qam}{QAM}{Quadrature Amplitude Modulation}

\newacronym{ntn}{NTN}{Non-Terrestrial Networks}
\newacronym{mdelan}{MDELAN}{Multi-Dilated Efficient Layer Aggregation Networks}
\newacronym{leo}{LEO}{Low Earth Orbit}
\newacronym{elan}{ELAN}{Efficient Layer Aggregation Networks}

\author{%
Mahdi Abdollahpour
\quad
Bruno De Filippo
\quad
Carla Amatetti
\quad
Alessandro Vanelli-Coralli
\\
{\small
 Department of Electrical, Electronic, and Information Engineering, University of Bologna, 40136 Bologna, Italy\quad
}
\\
{\small\itshape%
\{mahdi.abdollahpour, bruno.defilippo,  carla.amatetti2, alessandro.vanelli\}@unibo.it
}
}

\begin{document}
\title{On-board AI-based Channel Estimation for LEO NTNs \\
}
\maketitle

\begin{abstract}
Artificial Intelligence(AI) methods have shown strong channel estimation performance in terrestrial networks, but they typically rely on substantial computational resources. 
As 6G moves toward a unified architecture that will include Non-Terrestrial Networks (NTN) from day 0, availability of large and power hungry computational resources shall not be taken for granted.
At the same time, NTN propagation often exhibits high predictability, limited multipath richness and significant Doppler shifts, representing a specific channel estimation problem.
In this work, we propose a lightweight convolution-based channel estimator designed specifically for NTN operation and real-time onboard inference. 
We evaluate its channel estimation accuracy under stringent NGSO power budgets and quantify the resulting end-to-end impact on link performance. 
We show the improvement in terms of Mean Squared Error (MSE) achieved by the proposed approach compared with established algorithms, demonstrating that efficient AI models can deliver robust performance even on power-constrained spaceborne nodes. 
In addition, the proposed design by exploiting the domain knowledge, improves parameter efficiency by $27\%$ compared with state-of-the-art AI models and requires approximately $29\times$ fewer floating-point operations than conventional methods while achieving superior MSE performance.
\end{abstract}


\section{Introduction}

\gls{ai}-based physical-layer processing has recently emerged as a promising approach for improving wireless receiver performance. 
In particular, neural receivers and neural channel estimators have demonstrated clear gains over conventional model-based methods in several terrestrial \gls{5g} scenarios~\cite{wiesmayr2025design,honkala2021deeprx, honkala2026eqdeeprx}. 
However, many of these gains have been achieved with architectures that require substantial computational and memory resources, which limits their practicality in deployment-constrained platforms.

This challenge becomes even more critical for \gls{ntn}. 
As \gls{3gpp} and the broader \gls{6g} roadmap move toward unified terrestrial and non-terrestrial connectivity, \gls{leo} satellite systems are expected to support direct access with standard-compliant \gls{ofdm}-based waveforms~\cite{3gpp_ntn_overview, itu_m2160}. 
In such systems, receiver-side processing must operate under stringent power, memory, and onboard computing constraints. At the same time, \gls{ntn} propagation differs fundamentally from terrestrial channels due to large Doppler shifts, long propagation delays, limited multipath richness, and scenario-dependent channel statistics~\cite{guidotti2022path}. 
These characteristics create both new challenges and new opportunities for AI-based channel estimation.

\gls{ntn}-specific learning-based physical-layer processing has started to emerge only recently. In particular, \gls{ntn} channel estimation has been studied in settings such as \gls{leo} channel prediction and recent \gls{ai}-based estimators for \gls{leo} \gls{ntn} channels~\cite{amatetti2026channel, mahboob2024revolutionizing, pawase2026enhanced} have been proposed. These results suggest that AI can effectively exploit the structured dynamics of satellite channels. However, existing \gls{ntn}-oriented learning approaches are typically developed for a narrow problem setup or rely on complex models, and do not explicitly address the stringent compute and memory constraints of real-time onboard inference. This motivates the design of lightweight \gls{ai}-based channel estimators tailored to realistic \gls{leo} \gls{ntn} deployments.

Recent neural receiver designs have shown that low-complexity, model-driven architectures can substantially reduce the computational burden of \gls{ai}-enhanced baseband processing while preserving strong performance. 
In our previous work~\cite{mdx2025}, we introduced the \gls{mdx} receiver as a lightweight neural architecture for \gls{5g} \gls{nr} \gls{pusch} processing. 
While that design was motivated by edge-RAN deployment, its low-complexity structure also makes it a strong candidate for \gls{ntn} operation, where efficient inference is even more important. 

In this paper, we investigate AI-based channel estimation for \gls{6g} \gls{pusch} in \gls{leo} \gls{ntn} scenarios using realistic channel data generated with the QuaDRiGa \gls{ntn} extension~\cite{jaeckel2014quadriga}. 
Building on the \gls{mdx} design, we further introduce a new \gls{ai} block, termed \gls{mdelan}, to improve parameter efficiency and channel estimation accuracy while maintaining low complexity. 
We evaluate the proposed models in rural, suburban, and urban \gls{ntn} environments and compare them against practical conventional baselines.

The main contributions of this work are summarized as follows:
\begin{itemize}
    \item We evaluate the channel estimation performance of the low-complexity \gls{mdx} architecture in realistic \gls{leo} \gls{ntn} rural, suburban, and urban scenarios.
    \item We introduce a new lightweight neural block, \gls{mdelan}, to improve the efficiency and performance of \gls{ai}-based channel estimation.
    \item We assess \gls{pusch} channel estimation performance in terms of \gls{mse} using realistic \gls{ntn} channel realizations generated with QuaDRiGa.
    \item We compare the proposed neural models with conventional \emph{LS} and \emph{LMMSE} baselines and provide a computational complexity analysis in terms of \glspl{mac}, \glspl{flop}, and learnable parameters.
\end{itemize}


\section{System Model}



\begin{figure}[tbp] 
\centerline{\includegraphics[width=.9\columnwidth]{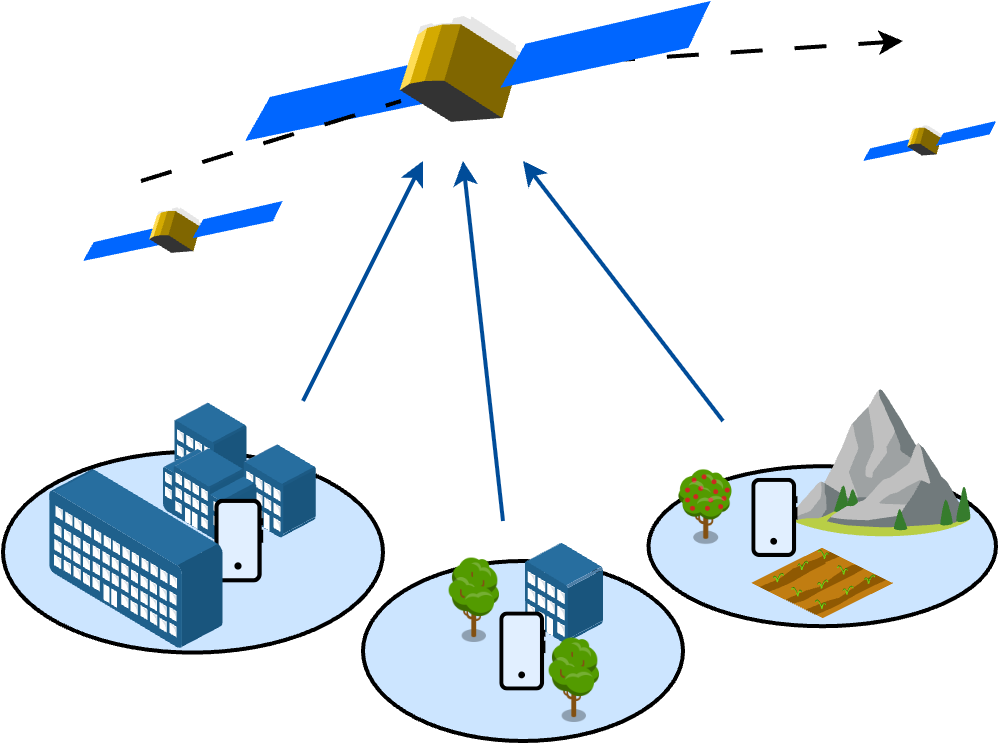}}
\caption{\gls{6g} NTN uplink system model with direct \gls{ue}-to-\gls{leo} satellite transmission in urban, suburban, and rural environments.}
\label{fig:sys}
\end{figure}
We consider a \gls{6g} \gls{pusch} transmission directly from a handheld \gls{ue} to a \gls{leo} satellite in S-band (2GHz) with a regenerative payload as illustrated in Fig.~\ref{fig:sys}. The \gls{ue} is equipped with a dual polarized omni directional antenna described in ~\cite[Table 6.1.1.1-3]{3gpp_tr_38821}. The satellite has parabolic reflector with dual-polarized two-port antenna described as \emph{LEO600} in ~\cite[Table 6.1.1.1-2]{3gpp_tr_38821}. The \gls{ue} is synchronized with the spaceborne gNodeB, with the residual Doppler assumed to be negligible with respect to the subcarrier spacing. The transmission consists of a \gls{6g} \gls{ofdm} frame structure with $S$ \gls{ofdm} symbols and carrying QPSK symbols over $F$ subcarriers. Pilot symbols are allocated in \gls{ofdm} symbols 2, and 11. The uplink received signal for each subcarrier $f\in\{1,\dots,F\}$, and for each \gls{ofdm} symbol $s\in\{1,\dots,S\}$ can be written as
\begin{equation}
    \mathbf{y}_{f,s}=\mathbf{h}_{f,s}x_{f,s}+\mathbf{n}_{f,s},
    \label{eq:system_model}
\end{equation}
where $\mathbf{y}_{f,s} \in \mathcal{C}^{2\times 1}$ is the received signal, $\mathbf{h}_{f,s} \in  \mathcal{C}^{2\times 1}$ is the channel vector, $x_{f,s} \in \mathcal{C}$ is the transmitted symbol, and $\mathbf{n}_{f,s} \in \mathcal{C}^{2}$ is the complex additive Gaussian noise with power spectral density $N_0$, distributed as $\mathcal{CN}(\mathbf{0},N_0\mathbf{I})$, with $\mathbf{I}$ representing the $2 \times 2$ identity matrix. The subscripts will be dropped for simplicity.

\subsection{Baseline Channel Estimators}
\label{sec:sys-baselines}
Based on the considered \gls{ofdm} transmission model, we use the following two conventional channel estimation methods as baselines:

\begin{itemize}
    \item \gls{ls}: this estimator at pilot locations computes the channel response independently on each pilot \gls{re} by dividing the received pilot symbols by the known transmitted pilots. Since this estimate is available only at pilot positions, interpolation is required to recover the channel over the entire \gls{ofdm} resource grid. A common low-complexity approach is linear interpolation, to estimates the channel at data \glspl{re}. This method is simple, and practical for real-time systems due to its low computational cost. However, its accuracy degrades when the channel varies rapidly or exhibits strong frequency selectivity, since linear interpolation cannot fully capture complex channel fluctuations.

    \item \gls{lmmse}: in contrast, \gls{lmmse} interpolation exploits second-order channel statistics, channel correlations in time and frequency, together with the noise variance, to obtain a minimum-mean-square-error estimate over the full grid. Compared with \gls{ls} followed by linear interpolation, \gls{lmmse} generally provides significantly better estimation accuracy, especially in highly selective channels. Nevertheless, the improvement comes at the cost of substantially higher computational complexity and the need for prior statistical knowledge or covariance estimation. Therefore, \gls{lmmse} interpolation is often less practical~\cite{savaux2017lmmse}.
\end{itemize}

\section{Model Driven Neural Receiver (MDX)}

\begin{figure*}[htbp] 
\centerline{\includegraphics[width=1\textwidth]{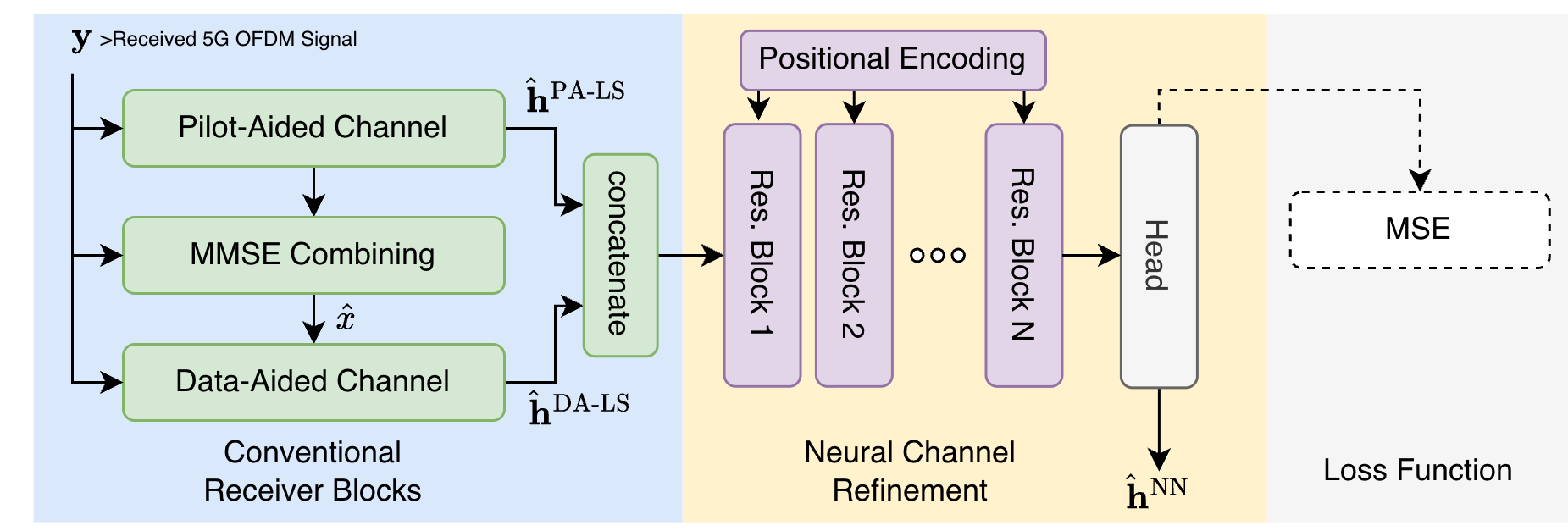}}
\caption{MDX channel estimation block diagram}
\label{fig:framework}
\vspace{-10pt} 
\end{figure*}

In this section, we describe \gls{mdx}~\cite{mdx2025} architecture configured for \gls{ntn} channel estimation. Additionally, we introduce new neural blocks to increase efficiency and performance of the receiver to accomodate the constraints of the satellite payload. The overall architecture of the channel estimation \gls{mdx} is shown in Fig.~\ref{fig:framework}. The model consists of conventional receiver blocks to obtain \gls{pals} and \gls{dals} channel estimates. Then a neural network with residual connections consisting of $N$ \glspl{resblock} is used to fuse the two kinds of estimates.

The \gls{pals} channel estimate is obtained at pilot \glspl{re}, and can be written as
\begin{equation}    
\hat{\mathbf{h}}=\frac{p^* \mathbf{y}}{|p|^2},
\label{eq:pa_ls}
\end{equation}
where $\hat{\mathbf{h}}^\text{PA-LS} \in \mathcal{C}^{2}$, \( p\) denotes the complex-valued pilot symbol, and \( \mathbf{y} \) is the received signal at the corresponding \gls{re}. 
Then $\hat{\mathbf{h}}^\text{PA-LS}$ is linearly interpolated into the \gls{ofdm} frame to obtain channel estimates for all \glspl{re}.

Then \gls{lmmse} combination method is used to estimate the transmitted symbols as
\begin{equation}    
\hat{x}=\left( \hat{\mathbf{h}}^H \hat{\mathbf{h}}+\hat{\sigma}^2 \right)^{-1} \hat{\mathbf{h}}^H \mathbf{y},  
\label{eq:mmse_comb}
\end{equation}
where $\hat{\sigma}$ is the estimated noise variance. 

Using the combined signal $\hat{x}$ we estimate \gls{dals} channel as 
\begin{equation}
\tilde{\mathbf{h}}  =  
 \mathbf{y} \cdot \hat{x}^*.
\label{eq:da_ls}
\end{equation}

\subsection{Channel Refinement}

The \gls{pals} and \gls{dals} channel estimates are processed further with a neural network to get a refined estimate of the channel. The network consists of Resnet style blocks with residual connections. The residual connections have a per-\gls{prb} learnable multiplier. For more details please refer to~\cite{mdx2025}. We use normalized horizontal and vertical location of every \gls{re} as \gls{pe},
\[
\mathcal{P} \in \mathbb{R}^{F \times S \times 2},
\]
where for each resource element \((f,s)\), the two channels correspond to the normalized vertical and horizontal locations:
\[
\mathcal{P}_{f,s,:} =
\begin{bmatrix}
\mathcal{P}_{f,s,1} \\
\mathcal{P}_{f,s,2}
\end{bmatrix}.
\]


The vertical location is normalized with respect to a block of $N_{\rm PRB}^{\rm ref}=3$ \glspl{prb} (all \glspl{prb} comprising $N_{SC}^{\mathrm {PRB}}=12$ subcarriers) along the subcarrier axis, and the horizontal location is normalized over the span of the OFDM symbol axis.

\begin{equation}
\mathcal{P}_{f,s,1} = \frac{f-1}{N_{\rm PRB}^{\rm ref} \cdot N_{SC}^{\mathrm {PRB}}},
\qquad
\mathcal{P}_{f,s,2} = \frac{s-1}{S}.
\label{eq:pe}
\end{equation}

The input to the neural network comprises:
\begin{enumerate}
\item The \gls{pals} channel estimate, transformed to a real tensor of shape $2 \times F \times S \times 2$ (the last dimension presents the real and imaginary part of the corresponding channel estimates, the first dimension represents the receiver antenna ports); 

\item The \gls{dals} channel estimate, similarly transformed to a real tensor of shape $2 \times F \times S \times 2$; and

\item The positional encoding tensor as defined in~\eqref{eq:pe}.
\end{enumerate}

These tensors are concatenated along the last dimension to get a tensor with 6 channels and spatial dimensions of $F \times S$. 

\subsection{Multi-Dilated Efficient Layer Aggregation Networks}

\begin{figure}[tbp] 
\centerline{\includegraphics[width=1\columnwidth]{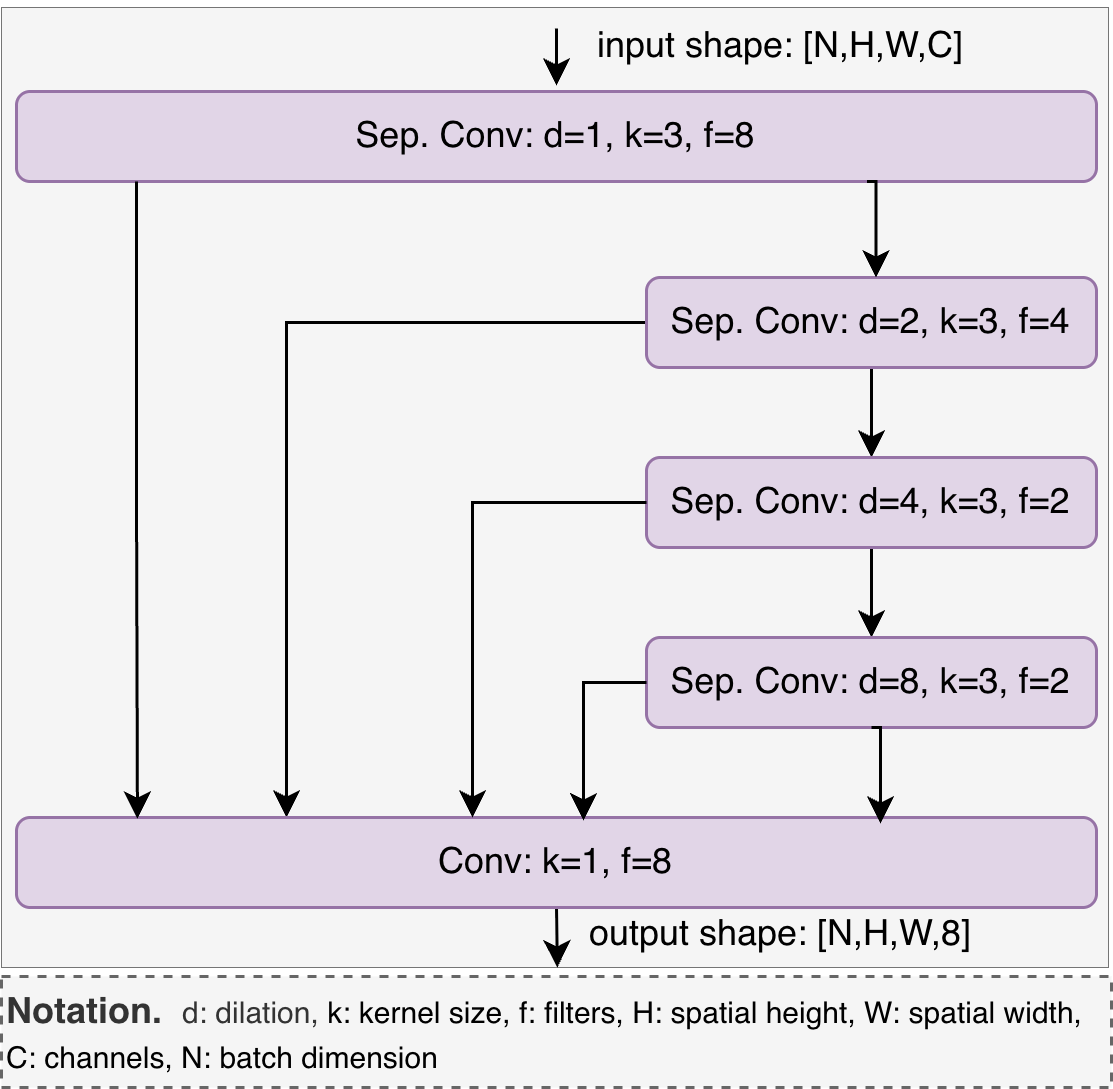}}
\vspace{-8pt} 
\caption{\gls{mdelan} Block Diagram}
\label{fig:mdelan}
\vspace{-10pt} 
\end{figure}

We introduce a new block, inspired by \gls{elan}~\cite{wang2022designing, wang2023yolov7}, illustrated in Fig.~\ref{fig:mdelan}.
The \gls{elan} is an architectural design strategy introduced to enhance the learning capability of deep neural networks by optimizing gradient propagation paths. 
Unlike traditional architectures where increasing depth can lead to the lengthening of the shortest gradient path, \gls{elan} controls the shortest and longest gradient paths, allowing the network to learn more diverse features and achieve higher parameter utilization efficiency.
We use multiple progressive dilations in our \gls{mdelan} structure to capture the spectral and temporal correlations of the \gls{ofdm} channel structure, increasing the receptive field without enlarging the kernels. 
To lower the complexity we use depthwise separable convolutions~\cite{chollet2017xception} throughout the structure.

\subsection{Neural Model}

The neural model is the same as introduced in~\cite{mdx2025}, with residual connections and per-\gls{prb} residual learnable multipliers, producing the real and imaginary components of the channel estimate (Fig.~\ref{fig:framework}). The head block comprises a depthwise separable convolution with two filters and no activation function. We here consider two model variants:
\begin{enumerate}
    \item In the first variant, the residual blocks are implemented using depthwise separable convolutions, following~\cite{mdx2025}. This model, denoted as \emph{MDX}, consists of five \glspl{resblock}.
    \item In the second variant, the residual blocks are implemented by the proposed \gls{mdelan} blocks. This model, denoted as \emph{MDX: MDELAN}, uses two \glspl{resblock}.
\end{enumerate}

Both variants are designed to have comparable computational complexity, enabling a fair comparison of their performance.

\subsection{Training}




The model is trained using the average \gls{mse} loss on a batch of size $N_{\text{TTI}}$, with each sample corresponding to a \gls{tti}:
\begin{equation}
    \mathcal{J} = \frac{1}{N_{\text{TTI}}} \sum_{n_{\text{TTI}}=1}^{N_{\text{TTI}}} \frac{1}{2FS} \sum_{s,f}
    \left\| \hat{\mathbf{h}}^{\mathrm{NN}}_{s,f} - \mathbf{h}_{s,f} \right\|_F^2,
\end{equation}
where $\hat{\mathbf{h}}^{\mathrm{NN}}$ denotes the neural network estimate, $\mathbf{h}$ is the ground-truth channel, and $\|\cdot\|_F$ is the Frobenius norm.
\section{Simulation Results}

In this section, we evaluate the proposed neural models in realistic \gls{ntn} scenarios and compare them with the conventional \emph{LS} and \emph{LMMSE} baselines introduced in Section~\ref{sec:sys-baselines}.

\subsection{Simulation Setup}

We train the neural models using the Adam optimizer with a learning rate of $10^{-3}$ and a regularization parameter of $0.01$. The end-to-end \gls{pusch} \gls{ofdm} transmission chain is simulated in Sionna~\cite{hoydis2022sionna}. The \gls{ntn} training and test channel data are generated using the QuaDRiGa channel simulator~\cite{jaeckel2014quadriga}. For each sample, a user is randomly dropped within the beam footprint of a \gls{leo} satellite. The user speed is drawn from a uniform distribution over $[0,100]$\,km/h. Further simulation details are summarized in Table~\ref{tab:data_config}. The neural models were trained on an NVIDIA GeForce GTX 1080 Ti GPU for 10,000 iterations with a batch size of $N_{\text{TTI}} = 128$.

\begin{table}[tbp]
\caption{Training and test configuration.}
\label{tab:data_config}
\centering
\begin{tabular}{|l|c|c|}
\hline
\textbf{Parameter} & \textbf{Train} & \textbf{Test} \\ \hline
Carrier frequency & \multicolumn{2}{c|}{2\,GHz} \\ \hline
Channel model & NTN dense urban, urban & rural, suburban \\ \hline
Speed & \multicolumn{2}{c|}{Uniformly distributed in $[0,100]$ km/h} \\ \hline
SNR & uniform $[-2,15]$ dB & $[-10,10]$ dB \\ \hline
PRBs & 5 & 11 \\ \hline
Subcarrier spacing & \multicolumn{2}{c|}{30\,kHz} \\ \hline
TTI length & \multicolumn{2}{c|}{14 OFDM symbols: 500\,$\mu$s} \\ \hline
Modulation & \multicolumn{2}{c|}{QPSK} \\ \hline
Pilot symbol indices & \multicolumn{2}{c|}{ \{2, 11\} (zero-based)} \\ \hline
\end{tabular}
\end{table}

\subsection{Rural Scenario}

\begin{figure}[tbp]
\centerline{\includegraphics[width=1\columnwidth]{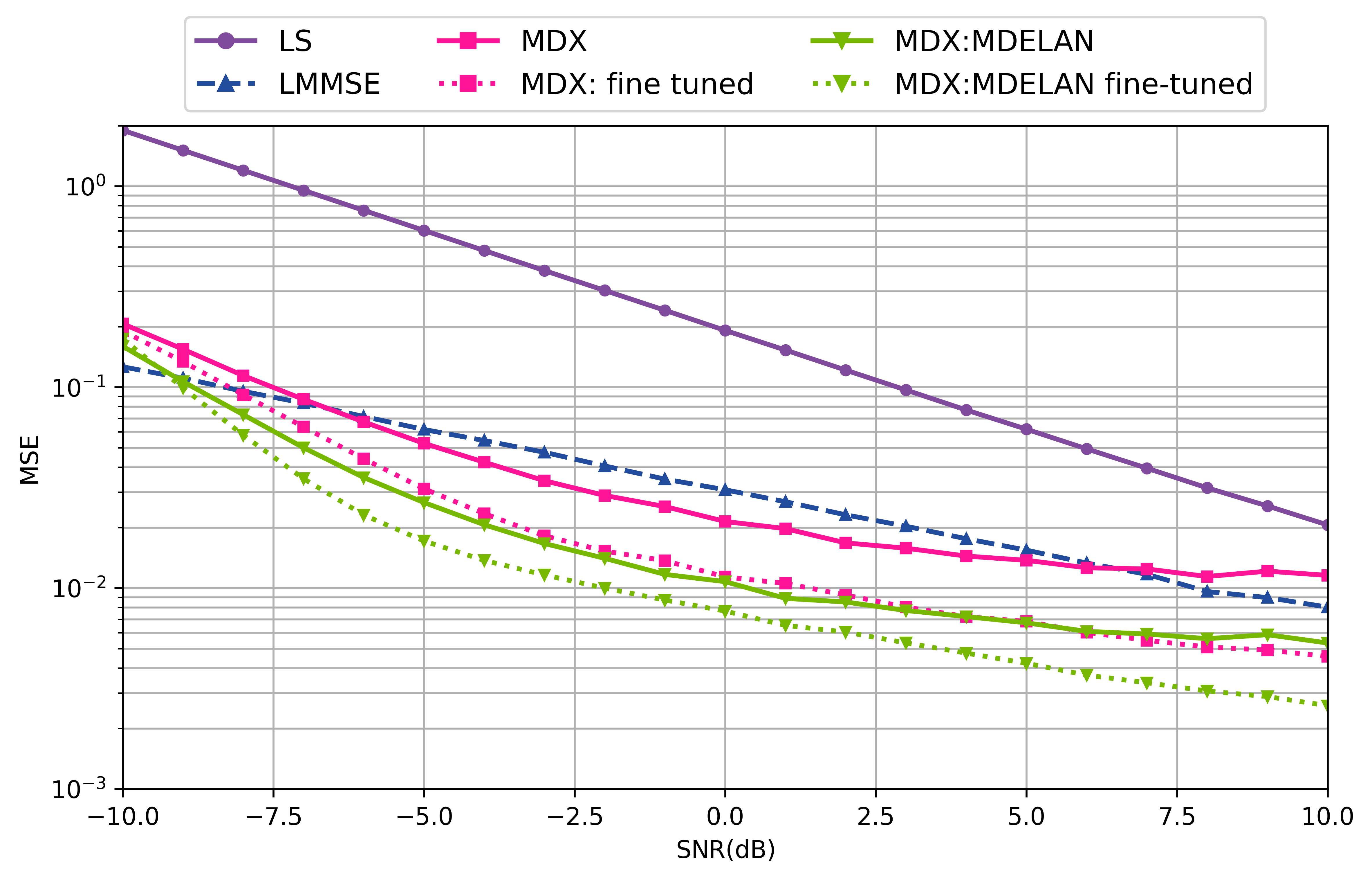}}
\caption{\gls{mse} vs. \gls{snr} performance in \gls{ntn} rural scenario. }
\label{fig:rural}
\end{figure}

Fig.~\ref{fig:rural} depicts the \gls{mse} as a function of \gls{snr} in the \gls{ntn} rural scenario. All neural approaches clearly outperform the practical \emph{LS} baseline over the entire \gls{snr} range. Without scenario-specific adaptation, \emph{MDX} achieves performance on the same order as \emph{LMMSE}, while the proposed \emph{MDX:MDELAN} further reduces the \gls{mse} and is the strongest non-fine-tuned neural model across the evaluated range. Fine-tuning improves both neural models. In particular, the fine-tuned \emph{MDX:MDELAN} attains the lowest \gls{mse} among all considered methods and maintains a clear advantage over \emph{MDX}. Additionally, the fine-tuned \emph{MDX} reaches performance close to that of the non-fine-tuned \emph{MDX:MDELAN}, further highlighting the stronger generalization capability of the proposed \emph{MDX:MDELAN} to the unseen rural scenario.


\subsection{Suburban Scenario}


\begin{figure}[t]
\centerline{\includegraphics[width=1\columnwidth]{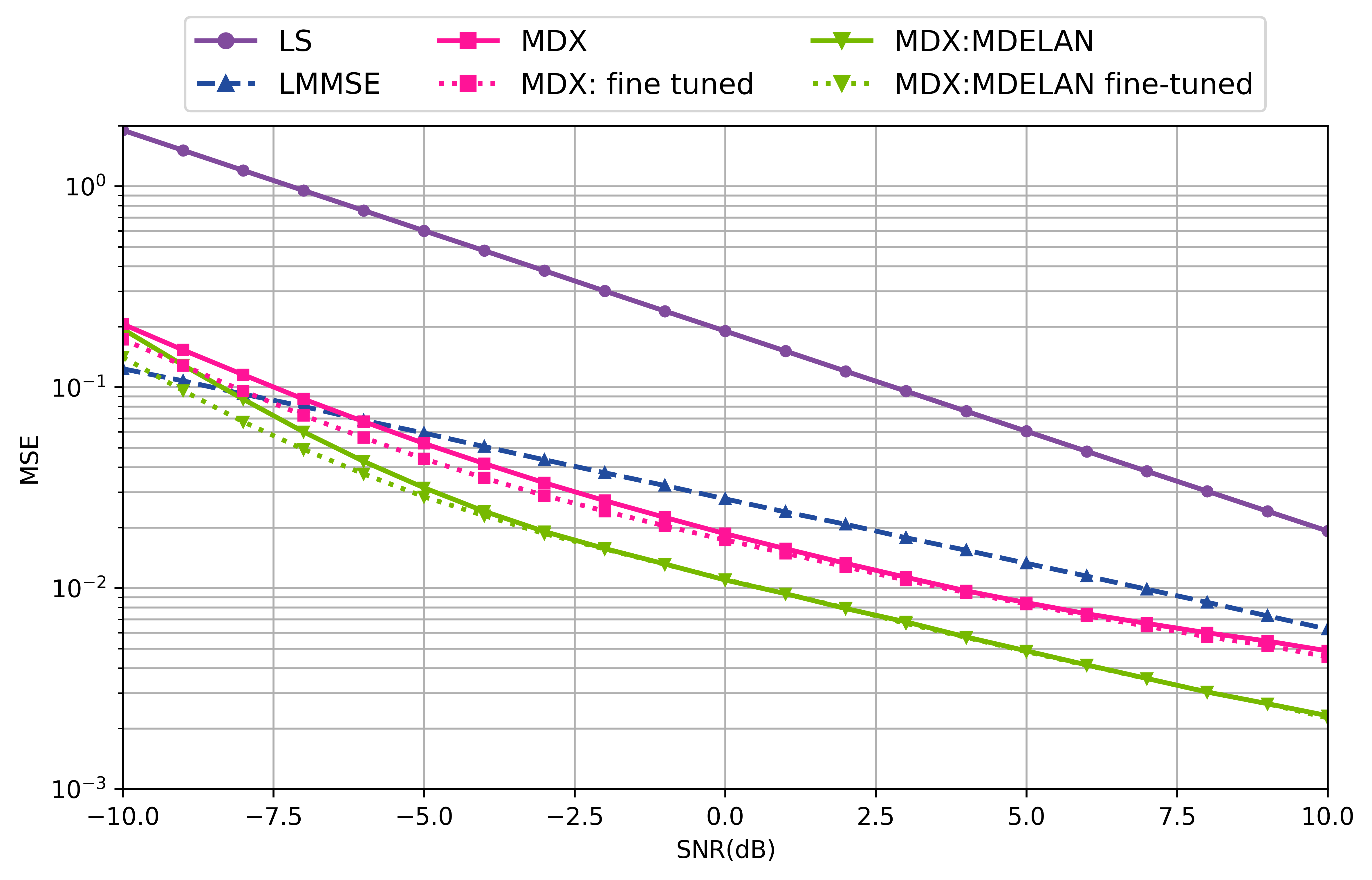}}
\caption{\gls{mse} vs. \gls{snr} performance in \gls{ntn} suburban scenario.}
\label{fig:suburban}
\end{figure}

Fig.~\ref{fig:suburban} reports the \gls{mse} performance in the \gls{ntn} suburban scenario. Similar to the rural case, all neural models significantly outperform the \emph{LS} baseline across the entire \gls{snr} range. \emph{MDX} achieves lower \gls{mse} than \emph{LMMSE}, while the proposed \emph{MDX:MDELAN} outperforms both \emph{MDX} and \emph{LMMSE}, with more pronounced gains at high \glspl{snr}. Fine-tuning on the suburban scenario provides only marginal improvement for neural models. This can be attributed to the similarity between the suburban environment and the training scenarios (urban and dense urban), suggesting that the models already generalize well without additional adaptation.

\subsection{Complexity Analysis}
All proposed neural blocks employ depthwise separable convolutions to reduce computational complexity. 
As a result, the overall complexity of the models is dominated by these operations. A depthwise separable convolution consists of a depthwise convolution followed by a pointwise ($1\times1$) convolution. 
Assuming the input tensor have spatial dimensions $F \times S$ (corresponding to the \gls{ofdm} grid), with $C_{\mathrm{in}}$ input channels and $C_{\mathrm{out}}$ output channels, for a kernel size of $k=3$, the total number of \glspl{mac} is given by
\vspace{-3pt} 
\begin{equation}
\mathrm{MACs} = 9FS C_{\mathrm{in}} + FS C_{\mathrm{in}}C_{\mathrm{out}},
\label{eq:macs_cov}
\vspace{-5pt} 
\end{equation}

\begin{figure}[tbp]
\centerline{\includegraphics[width=1\columnwidth]{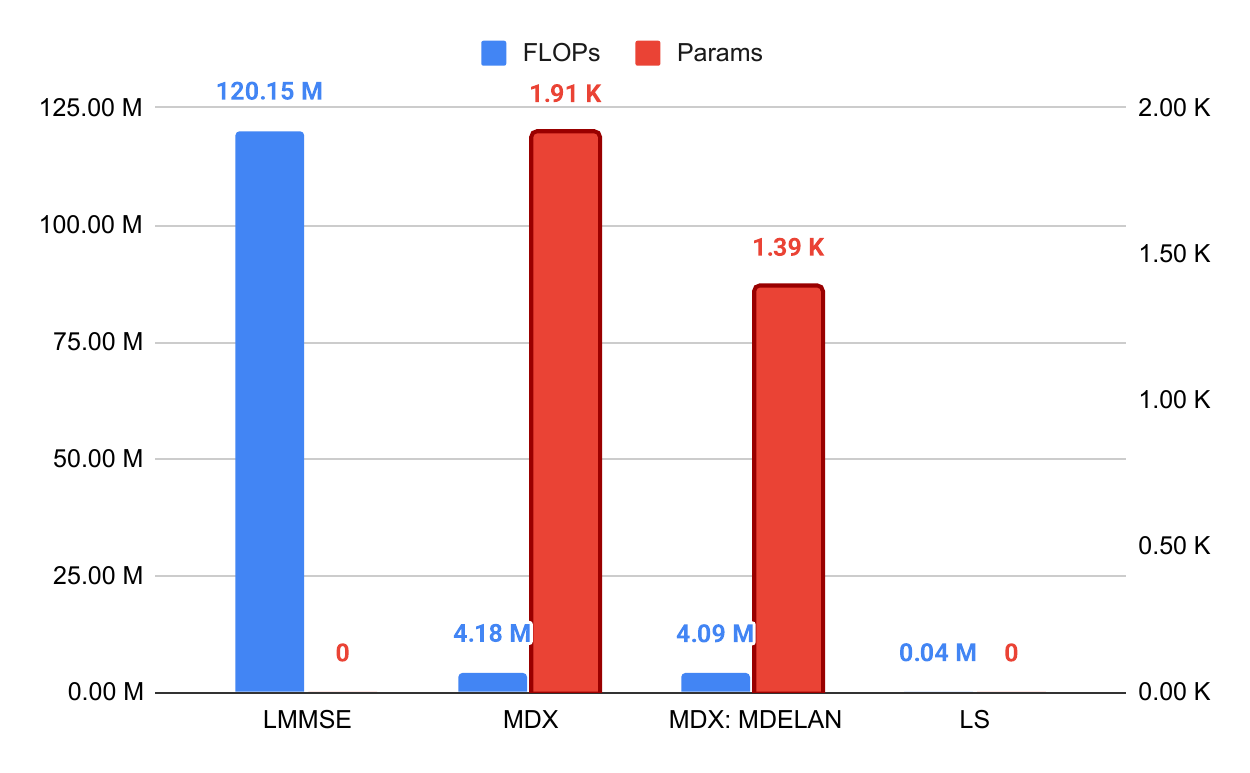}}
\caption{\glspl{flop} and number of parameters}
\label{fig:flops}
\end{figure}

The number of \glspl{flop} is approximated as $\mathrm{FLOPs} \approx 2 \times \mathrm{MACs}$. Each \gls{mdelan} block requires approximately $933$\,K \glspl{mac}, while each \gls{mdx} \gls{resblock} requires about $382$\,K \glspl{mac}. The overall computational complexity of the proposed models and the conventional baselines in the test configuration used in this paper is summarized in Fig.~\ref{fig:flops}. The proposed neural models achieve a significant reduction in computational complexity compared to the \emph{LMMSE} method, requiring approximately $29\times$ fewer \glspl{flop} while achieving higher estimation accuracy. As expected, the \emph{LS} method remains the least complex baseline at the expenses of the estimation MSE. Furthermore, the proposed \gls{mdelan} block demonstrates improved parameter efficiency compared to \gls{mdx}, achieving a reduction of approximately $27\%$ in the number of learnable parameters while maintaining superior performance.


\section{Conclusions}

This paper presented neural network-based channel estimation methods for \gls{6g} \gls{pusch} transmissions in \gls{ntn} \gls{leo} satellite scenarios under realistic channel models. The proposed \gls{ai}-based methods consistently outperform the practical \emph{LS} baseline and achieve performance comparable to, or better than, \emph{LMMSE}. In addition, they require approximately $29\times$ fewer \glspl{flop} than \emph{LMMSE} and do not rely on second-order channel statistics, making them both computationally efficient and practical for deployment. Furthermore, the proposed \gls{mdelan} block improves the parameter efficiency of the previous model by $27\%$ while also delivering better channel estimation performance. Future works will focus on achieving additional complexity reductions, \textit{e.g.}, by further optimizing the network architecture, and introducing an early exit mechanism to reduce the number of \glspl{resblock} based on the channel conditions. To support reproducibility, the implementation of \emph{MDELAN} will be released as open source in our repository~\cite{open_source}.

\section*{Acknowledgments}
This work was supported by UNITY-6G project, which received funding from the Smart Networks and Services Joint Undertaking (SNS JU) under the European Union’s Horizon Europe research and innovation programme under Grant Agreement No 101192650. The views expressed are those of the authors and do not necessarily represent the project. The Commission is not liable for any use that may be made of any of the information contained therein.


\Urlmuskip=0mu plus 1mu\relax
\def\UrlBreaks{\do\/\do-}
\bibliographystyle{IEEEtran}
\bibliography{bib_old} 

\end{document}